\documentclass[amsmath,twocolumn,superscriptaddress,prl,aps]{revtex4}
\usepackage{amsthm,amsfonts,graphicx,verbatim,color}
\newcommand{\be}{\begin{equation}}
\newcommand{\ee}{\end{equation}}
\newcommand{\bea}{\begin{eqnarray}}
\newcommand{\eea}{\end{eqnarray}}

\newcommand{\lp}{\left(}
\newcommand{\rp}{\right)}

\newcommand{\Tr}{{\rm Tr}\,}
\renewcommand{\phi}{\varphi}
\renewcommand{\epsilon}{\varepsilon}
\renewcommand{\vec}[1]{{\bf #1}}

\renewcommand{\cite}[1]{[\onlinecite{#1}]}
\newcommand{\diag}{\mathrm{diag}}

\begin{document}

\title{Chiral superconductivity from repulsive interactions in  doped graphene}
\author{Rahul Nandkishore}
\author{L. S. Levitov}
\affiliation{Department of Physics, Massachusetts Institute of Technology, Cambridge, Massachusetts 02139, USA}
\author{A. V. Chubukov}
\affiliation{Department of Physics, University of Wisconsin-Madison,
 Madison, Wisconsin 53706, USA}
%\email{ Correspondence should be addressed to AC}
\begin{abstract}
{\bf We identify graphene as a system where chiral superconductivity can be realized. Chiral superconductivity involves a pairing gap that winds in phase around the Fermi surface, breaking time reversal symmetry. 
We consider a unique situation arising in graphene at a specific level of doping, where the density of states is singular, strongly enhancing the critical temperature $T_c$. At this doping level, the Fermi surface is nested, allowing superconductivity to emerge from repulsive electron-electron interactions. We show using a renormalization group method that superconductivity dominates over all competing orders for any choice of weak repulsive interactions. Superconductivity develops in a doubly degenerate, spin singlet channel, and a mean field calculation indicates that the superconductivity is of a chiral $d+id$ type. We therefore predict that doped graphene 
can provide experimental realization of spin-singlet chiral superconductivity.}
\end{abstract}
\maketitle
 Chiral superconductors feature pairing gaps that wind in phase  around the Fermi surface (FS) by multiples of $2\pi$, breaking time-reversal symmetry (TRS) and parity
 and exhibiting a wealth of fascinating properties \cite{Sigrist,Volovikold,Sachdev}. The search for experimental realizations of chiral superconductivity %, a holy grail of correlated electron physics, 
greatly intensified in the last few years with the advent of topological superconductivity~\cite{Kane,Qi,Cheng}. 
 Here we show that chiral superconductivity with a $d_{x^2-y^2} \pm i d_{xy}$ ($d+id$) gap structure can be realized in graphene monolayer, a system of choice of modern nanoscience \cite{DasSarma, CastroNeto}.
We demonstrate that when graphene is doped to the vicinity of a Van Hove singularity in the density of states (DOS), repulsive electron-electron interactions induce d-wave superconductivity. Our renormalization group analysis indicates that superconductivity dominates over competing density wave orders, and also indicates that interactions select the  chiral  $d+id$ state over TRS-preseving $d$-wave states. 
The nontrivial topology of the $d+id$ state \cite{Volovikold} manifests itself in exceptionally rich phenomenology, including %a charge Hall 
 %effect at zero magnetic field \cite{Volovikold}, 
 a quantized spin and thermal Hall conductance \cite{Horovitz}, and a quantized boundary current in magnetic field \cite{kapitulnik}.
%LL , which 
%LL \mpar{removed majorana reference}%as well as unconventional Majorana modes \cite{Sato,Mao},} 
%LL  make it a highly desirable state in diverse areas of nanoscience.

 \begin{figure}[h]
 \includegraphics[width = 0.99\columnwidth]{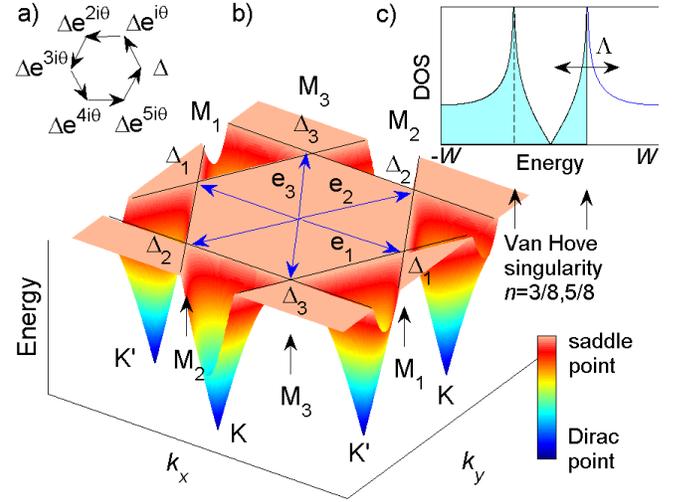} %{contours}
 \caption{\label{fig: dispersion}  Chiral superconductivity arises when graphene is doped to the Van Hove singularity at the saddle point ($M$ points of the Brillouin zone). a) $d+id$ pairing exhibiting phase winding around the hexagonal FS, which breaks TRS and parity ($\theta = 2\pi/3$). %The variation of the phase of the superconducting gap along the FS gives rise to persistent Josephson currents in momentum space, which manifestly break TRS.
  b) Conduction band for monolayer graphene \cite{CastroNeto}.
At $5/8$ filling of the $\pi$ band, the FS is hexagonal, and the DOS is logarithmically divergent (c) %(c). The divergent DOS  (c)
%  arises from 
at three inequivalent saddle points of the dispersion $M_{i}$ (i=1,2,3). Their location is given by $\pm \vec{e}_{i}$, where $2 \vec{e}_{i}$ is a reciprocal lattice vector.  
 The singular DOS strongly enhances the effect of interactions, driving the system into a chiral superconducting state (a). Since the FS is nested, superconductivity competes with density wave instabilities, and a full RG treatment is required to establish the dominance of superconductivity. %. The superconducting gap at the three saddle points takes values $(\Delta_1, \Delta_2, \Delta_3) = (\Delta, \Delta e^{2i\pi/3}, \Delta e^{-2i\pi/3}$)
 }
 \end{figure}
 
The search for chiral superconductivity has a long history.
 Spin-triplet p-wave chiral superconductivity ($p_x \pm i p_y$ state)
 has likely been  found in  $Sr_2RuO_4$ \cite{Mackenzie},  which represents a solid state analog of superfluid  ${}^3$He \cite{Volovikold}, but the \emph{spin-singlet} $d+id$ state has not yet been observed experimentally.
Such a state was once proposed as a candidate state for high $T_c$ cuprate superconductors \cite{kapitulnik, Horovitz}, but later gave way to
 a more-conventional TRS-preserving $d-$wave state. The key difficulty in realizing a $d+id$ state is that 
 the interactions that favor a d-wave state usually have strong momentum dependence and hence %usually exhibit a peak at a finite momentum transfer, comparable to inverse lattice spacing, and 
 %pairing interactions of this form 
 distinguish between $d_{x^2-y^2}$ and $d_{xy}$ pairing.  However, in graphene the $d_{x^2-y^2}$ and $d_{xy}$ pairing channels are degenerate by symmetry \cite{Gonzalez, Doniach},
%AC question -- did he said explicitly that d_{xy} and d_{x^2-y^2} are degenerate?
 opening the door to formation of a $d+id$ superconducting state.  
    
How can superconductivity be induced in graphene? 
 Existing proposals for superconductivity in undoped graphene rely on the conventional phonon mediated BCS mechanism \cite{Uchoa}, which leads to an $s-$wave superconductivity with low $T_c$ values for realistic carrier densities due to the vanishing density of states of relativistic particles.
 However, there is an alternative route to superconductivity, wherein repulsive microscopic interactions give rise to attraction in a d-wave channel \cite{Kohn}. This alternative route 
 becomes viable when graphene is doped to the $M$ point of the Brillouin zone
 corresponding to $3/8$ or $5/8$ filling of the $\pi$ band 
 (pristine graphene corresponds to $1/2$ filling). 
At this filling factor, a logarithmic Van Hove singularity originates from three inequivalent saddle points, and the FS also displays a high degree of nesting,
  forming a perfect hexagon when third and higher neighbor hopping effects are neglected \cite{CastroNeto, Gonzalez} (Fig.\ref{fig: dispersion}). The combination of a singular DOS and a near-nested FS 
strongly enhances the effect of interactions \cite{Furukawa, Shulz, Dzyaloshinskii}, allowing non-trivial phases to emerge at  relatively high  temperatures, even if interactions are weak compared to the fermionic bandwidth $W$.  
Relevant doping levels were recently achieved experimentally using calcium and potassium dopants \cite{McChesney}. Also, the new technique \cite{ionicliquid} which employs ionic liquids as gate dielectrics allows high levels of doping to be
reached without introducing chemical disorder.

{\it Competing orders:}
In systems with near-nested FS, superconductivity (SC) has to compete with
 charge density wave (CDW) and spin density wave (SDW) instabilities \cite{ricereview}.
  At the first glance, it may seem that a system with repulsive interactions should develop a density-wave order rather than become a superconductor. However, to analyze this properly, one needs to know the susceptibilities to the various orders at a relatively small energy, $E_0$, at which the order actually develops. The couplings at $E_0$  generally differ from their bare values because of renormalizations by fermions with energies between $E_0$ and $W$. At weak coupling, these renormalizations are well  captured by the renormalization group (RG) technique.
 
Interacting fermions with a nested FS and logarithmically divergent DOS have previously been studied 
on the square lattice using the RG methods\cite{Furukawa, Shulz, Dzyaloshinskii, ricereview}, where spin fluctuations were argued to stimulate superconciductivity.
However, analysis also revealed near degeneracy between SC and SDW orders. The competition between these orders is decided by subtle interplay between deviations from perfect nesting, which favor SC, and  subleading terms in the RG flow, which favor SDW.
In contrast, the RG procedure on the honeycomb lattice
 unambiguously selects SC at leading order, allowing us to safely neglect subleading terms. The difference arises because the honeycomb lattice contains three saddle points, whereas the square lattice has only two, and the extra saddle point tips the delicate balance seen on the square lattice between magnetism and SC decisively in favor of superconductivity. A similar tipping of a balance between SC and SDW in favor of SC has been found in RG studies of some Fe-pnictide superconductors \cite{chubukov, Thomale}.
 
In previous works on graphene at the $M$ point, various instabilities 
 were analyzed using the random-phase approximation (RPA) and mean field theory. In \cite{Gonzalez}, the instability to d-wave SC
 was studied, whereas \cite{Vozmediano} considered a  charge
 `Pomeranchuk' instability to a metallic phase breaking lattice rotation symmetry, and \cite{TaoLi, MoraisSmith} 
 considered a spin density wave (SDW)  
instability  to an insulating phase.
 Within the framework of mean field theory, utilized in the above works, all of these phases are legitimate 
  potential instabilities of the system. However, clearly graphene at the $M$ point cannot be simultaneously superconducting, metallic and insulating. The RG analysis treats all competing orders on an equal footing, and predicts that
 the {\it dominant} weak coupling instability is to superconductivity, for any choice of repulsive interactions, even for perfect nesting. 
Further, the Ginzburg-Landau theory constructed near the RG fixed point favors the $d+id$ state.

\begin{figure}[h]
\includegraphics[width = \columnwidth]{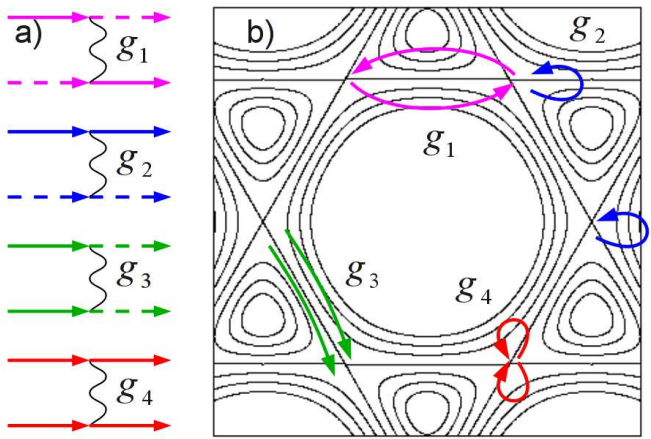} 
\caption{
Possible interactions in the patch model. (a) Feynman diagrams representing allowed two-particle scattering processes among different patches, Eq.\ref{eq: action}. Solid and dashed lines represent fermions on different patches, whereas wavy lines represent interactions. (b) Pictorial representation of these scattering processes, superimposed on a contour plot of the energy dispersion. %The processes (a,b) describe two particle scattering between three inequivalent saddle points. 
Each scattering process comes in three flavors, according to the patches involved. %- i.e., there is $g_1^{M_1M_2}$, $g_{1}^{M_1M_3}$ and $g_1^{M_2M_3}$, and likewise for $g_2, g_3, g_4$. 
However, it follows by symmetry that %$g_i^{M_1M_2}=g_i^{M_1M_3}=g_i^{M_2M_3}$, 
the scattering amplitudes are independent of the patches involved, and therefore we suppress the flavor labels. 
\label{fig: interactions}}
\end{figure}

  {\it The model:}  We follow the procedure developed for the square lattice \cite{ricereview} and construct a patch RG that considers only fermions near three saddle points, 
which dominate the DOS. There are four distinct interactions in the low energy theory, involving two-particle scattering between different patches, as shown in Fig.\ref{fig: interactions}.

The system is described by the low energy theory
\bea
&&\mathcal{L} = \sum_{\alpha=1}^3 \psi^{\dag}_{\alpha} (\partial_{\tau} - \epsilon_{\vec{k}} + \mu) \psi_{\alpha} - \frac12 g_4 \psi^{\dag}_{\alpha }\psi^{\dag}_{\alpha} \psi_{\alpha} \psi_{\alpha} \label{eq: action} 
\\\nonumber
&& - \sum_{\alpha \neq \beta} \frac12 \big[g_1 \psi^{\dag}_{\alpha }\psi^{\dag}_{\beta} \psi_{\alpha} \psi_{\beta} + g_2 \psi^{\dag}_{\alpha }\psi^{\dag}_{\beta} \psi_{\beta} \psi_{\alpha} + g_3\psi^{\dag}_{\alpha }\psi^{\dag}_{\alpha} \psi_{\beta} \psi_{\beta}\big]
,
\eea
where summation is over patch labels $\alpha, \beta = M_1, M_2, M_3$. Here $\epsilon_{\vec{k}}$ is the tight binding dispersion, expanded up to quadratic terms about 
each saddle point. For example, near point $M_1$, the tight-binding model \cite{Wallace} predicts dispersion 
$\epsilon_{\vec{k}} = 2 \pi^2 a^2 t \big((\delta k_x)^2 - \sqrt{3} \delta k_x \delta k_y + O((\delta k)^4)\big)$, where $t$ is the nearest neighbor hopping, and $a$ is the lattice constant, and $\delta \vec k=\vec k-\vec k_{M_1}$. 
The chemical potential value $\mu=0$ describes system doped exactly to the saddle point.
We note that while the existence of saddle points is a topological property of the FS and is robust to arbitrarily long range hopping, the FS nesting is spoilt by third and higher neighbor hopping effects \cite{CastroNeto, Gonzalez}. Inequivalent saddle points are connected by a nesting vector $Q_{\alpha \beta} = \vec{e}_{\alpha} - \vec{e}_{\beta}$ (Fig.\ref{fig: dispersion}).  A spin sum is implicit in the above expression, and the interactions are assumed to be spin independent. The short-range interaction model, used in our analysis, is expected to provide a good approximation under the conditions of metallic screening arising due to the states near the FS. We further assume that screening is insensitive to the level of doping relative to the $M$ point. While these assumptions introduce a large uncertainty into the bare values for the interactions, we will show that precise knowledge of these bare values is not required for determining the final state.

The patch structure of the interactions is restricted by momentum conservation,
which allows only the four interactions in (\ref{eq: action}). The Umklapp interaction $g_3$ is allowed, because it conserves momentum modulo a reciprocal lattice vector. All four interactions in (\ref{eq: action}) are marginal at tree level, but acquire logarithmic corrections in perturbation 
theory, which come from energy scales $E < \Lambda$, where $\Lambda \approx t$ is the energy scale at which higher order corrections to the dispersion become important. 

Logarithmic divergences in perturbation theory analysis indicate that the problem is well suited to study using RG.  
The building blocks of the RG are the susceptibilities in the particle-particle and particle-hole channels, $\Pi_{\rm pp}$ and $\Pi_{\rm ph}$, evaluated respectively at momentum transfer zero and at momentum transfer $Q_{\alpha \ne \beta}$ (Fig.\ref{fig: dispersion}). Similarly to \cite{ricereview}, we have
\bea\label{eq: divergences}
&&\Pi_{\rm pp}(0) = \frac{\nu_0}{4} \ln \frac{\Lambda}{\max{(T,\mu)}} \ln \frac{\Lambda}{T}, \quad
\\\nonumber
&& \Pi_{\rm ph}(Q_{\alpha \ne \beta}) = \frac{\nu_0}{4}  \ln \frac{\Lambda}{\max{(T,\mu)}} \ln \frac{\Lambda}{\max{(T,\mu, t_3)}}, 
%AC ph -> pp ?
\eea
and $\Pi_{\rm ph}(0), \Pi_{\rm pp}(Q_{\alpha \ne \beta}) = \nu_0 \ln \frac{\Lambda}{\max{(T,\mu)}}$, where $\Lambda$ is our UV cutoff (Fig.\ref{fig: dispersion}) and $T$ is the temperature. The single spin density of states at a saddle point is $\nu_0 \ln \frac{\Lambda}{\max{(T,\mu)}}$.
 The additional $\log$ factor in $\Pi_{\rm pp}(0)$ (Cooper channel) arises because $\epsilon_{\vec{k}} = \epsilon_{-\vec{k}}$, generic for any system with time reversal or inversion symmetry. In contrast, the additional $\log$ factor in $\Pi_{\rm ph}(Q_{\alpha \ne \beta})$ arises from nesting of the FS, and is cut in the IR by any term that spoils the nesting, such as  third neighbor hopping $t_3$ or doping $\mu$ \cite{Gonzalez}. We assume $\max(t_3,\mu) \ll \Lambda$, so $\Pi_{\rm ph} (Q_{\alpha \ne \beta})$ and $\Pi_{\rm pp}(0)$ are of the same order under RG. 

{\it RG equations:} The RG equations are obtained by 
extending the approach developed for the square lattice problem \cite{Furukawa} to the number of patches $n>2$. The number of patches matters only in diagrams with zero net momentum in fermion loops, since it is only there that we get summation over fermion flavors inside the loop. The only zero-momentum loop with a $\log^2$ divergence is in the Cooper channel. Moreover, only the $g_3$ interaction changes the patch label of a Cooper pair, therefore, the number of patches affects only diagrams where two $g_3$ interactions are combined in the Cooper channel. With logarithmic accuracy, using $y =  \Pi_{\rm pp}(\vec{k} = 0, E) = \frac{\nu_0}{4} \ln^2 \frac{\Lambda}{E}$ as the RG time, we obtain the $\beta$ functions 
 \begin{eqnarray}
\frac{dg_1}{dy} &=& 2 d_1 g_1 (g_2-g_1), \qquad
 \frac{dg_2}{dy} = d_1(g_2^2 + g_3^2),
\nonumber\\\label{eq: beta functions}
\frac{dg_3}{dy} &=& - (n-2) g_3^2 - 2 g_3g_4 + 2 d_1g_3(2g_2-g_1), \qquad
\\\nonumber
\frac{dg_4}{dy} &=& -(n-1)g_3^2 - g_4^2 .
\end{eqnarray}
Here $d_1(y)= d \Pi_{\rm ph}(Q)/dy \approx \Pi_{\rm ph}(Q)/\Pi_{\rm pp}(0)$ is the `nesting parameter' \cite{Furukawa, ricereview}. This quantity equals one in the perfectly nested limit. For non-perfect nesting, $d_1(y)$ has the asymptotic forms $d_1(y=0)=1$, $d_1(y\gg 1) = \ln|\Lambda/t_3|/\sqrt{y}$, and interpolates smoothly in between. Since the RG equations flow to strong coupling at a finite scale $y_c$, we treat $0<d_1(y_c)<1$ as a parameter in our analysis.

The $\beta$-functions, Eq.(\ref{eq: beta functions}), reproduce the two-patch RG from \cite{Furukawa} when we take $n = 2$, and neglect subleading $O(\log)$ divergent terms ($d_{2,3}(y)$ from \cite{Furukawa}), and also reproduce for $n=2$ the RG equations for the Fe-pnictides \cite{chubukov}. Graphene near the Van Hove singularity however is described by $n=3$. 

We note from inspection of (\ref{eq: beta functions}) that $g_1, g_2$ and $g_3$ must stay positive (repulsive) if they start out positive. This follows because the $\beta$ function for $g_2$ is positive definite, and the $\beta$ functions for $g_1$ and $g_3$ vanish as the respective couplings go to zero. However, $g_4$  decreases under RG and eventually changes sign and becomes negative. As we will see,
  $g_3-g_4$ becomes large and positive
under RG, driving an instability to a superconducting phase. However, the positive $g_3$ coupling penalizes s-wave superconductivity, so pairing occurs in a higher angular momentum (d-wave) channel.

\begin{figure}[h]
\includegraphics[width = \columnwidth]{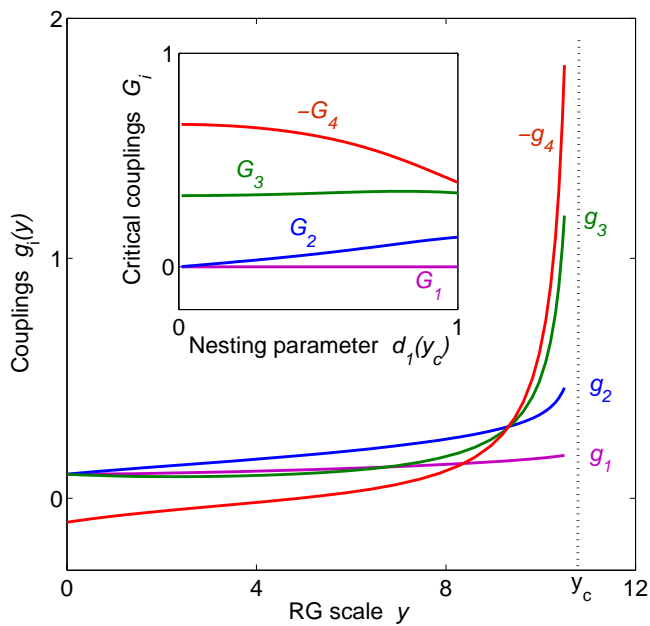}
\caption{Flow of couplings with RG scale $y$, starting from repulsive interactions. Note that the coupling $g_4$ changes sign and becomes attractive, leading to a (superconducting) instability at the energy scale $y_c$ (Eq.\ref{eq: tcsqrt}). Inset: Critical couplings $G_i$ (\ref{eq: critical couplings}) near $y_c$ as a function of the nesting parameter at the ordering energy scale, $d_1(y_c)$. The dominance of superconductivity over spin density wave order arises because $-G_4>G_2$ for all values of $d_1(y_c)$.   %
{\it Initial conditions:} The RG flow is obtained by numerical integration of (\ref{eq: beta functions}) with initial conditions $g_i(0) = 0.1$, and modeling the nesting parameter as $d_1(y) = 1/\sqrt{1+y}$. The qualitative features of the flow are insensitive to initial conditions, and to how we model $d_1$. The critical couplings (inset) are universal, and independent of initial conditions. %with the same critical couplings arising for any choice of initially repulsive interactions.  %Similar results are obtained if $d_1$ is treated as a constant.  
 \label{fig: rgintegration}}
\end{figure}

We integrate our RG equations with $n=3$ from starting from $g_i = g_0 = 0.1$ and modeling $d_1$ as $d_1(y) = 1/\sqrt{1+y}$. The results are plotted in Fig.\ref{fig: rgintegration}. Similar results are obtained if we 
 just treat $d_1$ as a constant. The couplings diverge at a scale $y_c \approx 1/g_0$, corresponding to a critical temperature and ordering energy scale
\begin{equation}
T_c,\,E_0 \sim \Lambda \exp(-A/\sqrt{g_0 \nu_0}) \label{eq: tcsqrt}.
\end{equation}
Here $A$ is a non-universal number that depends on how we model $d_1(y)$. For $d_1 = 1$ (perfect nesting, corresponding to zero third neighbor hopping $t_3$), we obtain $A = 1.5$. An RPA-type estimate of $g_0$ is outlined in the online supplementary material. While $T_c$ and $E_0$ are exponentially sensitive to $g_0$, thus introducing a considerable uncertainty to our estimate,
%LL In the online supplementary material, we use the random phase approximation (RPA) to estimate $g_0$. However, since $T_c$ is exponentially sensitive to $g_0$, our estimate comes with a very large uncertainty. Nevertheless, it is evident from (\ref{eq: tcsqrt}) that the characteristic energy scales for are strongly enhanced relative to the BCS result
a strong enhancement of characteristic energy scales relative to the BCS result is evident from Eq.(\ref{eq: tcsqrt}). 

 A similar $\sqrt{g_0}$ dependence arises in the treatment of color superconductivity \cite{Son}  and in the analysis of the pairing near quantum-critical points in 3D \cite{qcp}. It results in a $T_c$ that is strongly enhanced compared to the standard BCS result $T_c \sim \exp(-A'/g_0 \nu_0)$. It should be noted that the enhancement of $T_c$ in (\ref{eq: tcsqrt}) arises from weak coupling physics. %When $g_0$ becomes large, the separation of scales that allows us to neglect $O(\ln)$ terms disappears, and the increase of $T_c$ will likely be cut by $O(\ln)$ contributions from the vertex corrections and the self energy. In other words, in our analysis SC emerges from weak coupling physics.  
 It is distinct from the high $T_c$  superconductivity that could arise if the microscopic interactions were strong
 Refs.\cite{Baskaran, Doniach, Honerkamp, Roy}. % unconventional superconductivity was proposed, for undoped graphene, 
%  starting from strong correlations.
% However, since no such superconducting state has 
 %been observed
 %in undoped graphene, the strong coupling approach 
%does not appear to be suitable.

Returning to our RG analysis, we note that near the instability threshold, $g_1, g_2, g_3 \rightarrow \infty$ and $g_4 \rightarrow - \infty$, with $-g_4>g_3>g_2>g_1$. %
This observation may be made precise by noting that close to $y_c$, the interactions scale as 
\begin{eqnarray}
g_i(y) \approx \frac{G_i}{y_c-y} \label{eq: critical couplings}
\end{eqnarray}
Substituting into Eq.\ref{eq: beta functions}, we obtain a set of polynomial equations, which may be solved for the co-efficients $G_i$ as a function of $d_1(y_c)$. The solution is plotted in
 the inset of  Fig.\ref{fig: rgintegration}. 
  Note that $-G_4>G_3>G_2>G_1$ for all values of $d_1(y_c)$ satisfying $0\le d_1(y_c)\le1$.  We have verified that any choice of repulsive bare couplings leads to the same limiting trajectory (see online supplementary material).

\begin{figure}[h]
\includegraphics[width = \columnwidth]{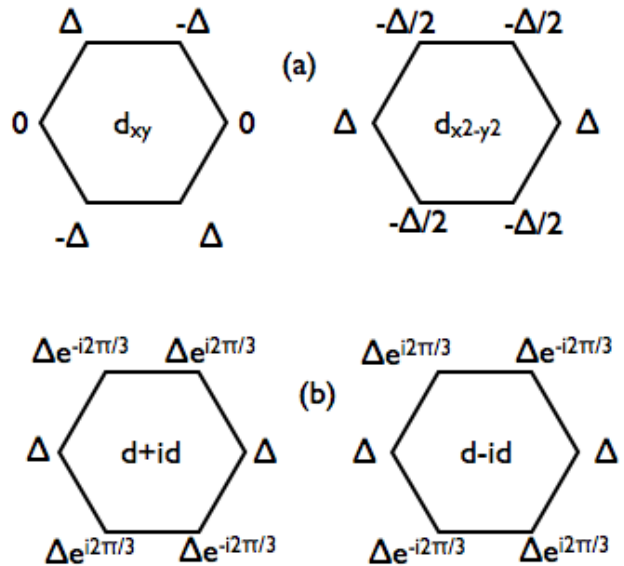}
\caption{Possible superconducting orders that could develop at the $M$ point. (a) A $d_{x^2 - y^2}$ or $d_{xy}$ state would be realised if $K_2<0$ in the Landau expression for the free energy, Eq.\ref{eq: landau} (b) The $d_{x^2-y^2}$ and $d_{xy}$ orders can co-exist if $K_2>0$ in Eq.\ref{eq: landau}. A microscopic calculation 
 indicates that the states (b) have lower free energy. \label{fig: orders}}
\end{figure}

{\it Susceptibilities:} We now investigate the instabilities of the system by evaluating the susceptibilities $\chi$ for various types of order. 
 To analyze the superconducting instability, we introduce infinitesimal test vertices corresponding to particle-particle pairing into the action, $\mathcal{L} = \mathcal{L}_0 + \delta \mathcal{L}$, where $\mathcal{L}_0$ is given by (\ref{eq: action}) and 
\begin{equation}
\delta \mathcal{L} = \sum_{\alpha = 1}^3 \tilde \Delta_{\alpha} \psi^{\dag}_{\alpha, \uparrow} \psi^{\dag}_{\alpha, \downarrow} +  \tilde \Delta^*_{\alpha} \psi_{\alpha, \uparrow} \psi_{\alpha, \downarrow}
,
\end{equation}
 one test vertex for each patch. The renormalisation of the test vertices
is governed by the equation \cite{Furukawa}
\begin{equation}
\label{eq: 3by3}
\frac{\partial}{\partial y} \left[ \begin{array}{c} \tilde \Delta_1 \\ \tilde \Delta_2 \\ \tilde \Delta_3 \end{array}\right] = - 2\left[\begin{array}{ccc} g_4 & g_3 & g_3 \\ g_3 & g_4 & g_3 \\ g_3 & g_3 & g_4\end{array} \right]\left[ \begin{array}{c} \tilde \Delta_1 \\ \tilde \Delta_2 \\ \tilde \Delta_3 \end{array}\right] 
\end{equation}
 which can be diagonalized by transforming to the eigenvector basis
\begin{eqnarray}
\tilde \Delta_a &=& \frac{\tilde \Delta}{\sqrt{2}}\big(0,1,-1\big), \quad  \tilde \Delta_b = \sqrt{\frac{2}{3}}\tilde \Delta \bigg(1,-\frac12,-\frac12\bigg) \label{eq: deltas} \\ \tilde \Delta_c &=& \frac{\tilde \Delta}{\sqrt{3}} \big(1,1,1\big)
. \label{eq: deltaswave}
\end{eqnarray}
Here $\tilde \Delta_c$ is an s-wave order, whereas  $\tilde \Delta_a $ and $\tilde \Delta_b$ correspond to order parameters that vary around the Fermi surface as $\tilde \Delta \cos(2\phi)$ and $\tilde \Delta \sin(2\phi)$, where $\phi$ is the angle to the $x$ axis (see Fig \ref{fig: orders}). 
Such dependence describes d-wave superconducting orders (SCd), since the gap changes sign four times along the FS. In 2D notation, the two order parameters $\tilde \Delta_a$ and $\tilde \Delta_b$ correspond to $d_{xy}$ and  $d_{x^2-y^2}$ superconducting orders respectively.
 
Notably, we find the s-wave vertex $\tilde \Delta_c$, Eq.(\ref{eq: deltaswave}),  has a negative  eigenvalue  
%AC ???....
 and 
 is suppressed under RG flow (\ref{eq: 3by3}). This is 
 to be expected given that we started out with repulsive microscopic interactions. At the same time, the d-wave orders $\tilde \Delta_a$ and $\tilde \Delta_b$ have (identical) eigenvalue $g_3 - g_4$, which 
 may be negative at the bare level but definitely 
becomes \emph{positive} under RG, indicating an instability in the d-wave channel. We solve (\ref{eq: 3by3}) for the d-wave orders by substituting the scaling form of the interactions (\ref{eq: critical couplings}), and find that 
 the d-wave susceptibility  diverges 
  near $y_c$ as 
\begin{equation}
\chi_{SCd}(y)=\frac{\tilde \Delta_{a,b}(y)}{\tilde \Delta_{a,b}(0)} \sim (y_c-y)^{2(G_4-G_3)}, %AC suggest to cut
%\sim \bigg(\frac{g_4(y)}{-g_0}\bigg)^{2(G_3-G_4)}, 
\label{eq: exponents}
\end{equation}
  where, we remind, $G_3 - G_4 >0$. 
 
The divergence of the SCd susceptibility indicates an instability to d wave superconductivity under RG, with the $\tilde \Delta \cos(2\phi)$ and $\tilde \Delta \sin(2\phi)$ order parameters having {\it identical} susceptibility. The degeneracy of the two d-wave orders is guaranteed, since the $d_{x^2 - y^2}$ and $d_{xy}$ functions belong to the same two dimensional irreducible representation of the lattice point group \cite{Gonzalez, Doniach}.  However, this does not guarantee that d-wave superconductivity will develop, since the SCd instability must compete against the tendency for density wave formation. 

To investigate density wave formation, we introduce test vertices representing pairing of particles with holes on a different patch. The particles and holes may
 pair in the charge channel, forming CDW, or in the spin channel, forming SDW.
% have the same spin (CDW) or opposite spin (SDW).
 We compute the renormalization of the pairing vertices under RG, and find that the CDW vertex is suppressed by interactions, but the SDW vertex $\tilde \Delta_{SDW}$ is enhanced, similar to \cite{Furukawa}. The SDW susceptibility $\chi_{SDW}$ diverges near $y_c$ as 
\begin{eqnarray}
\chi_{SDW}= \frac{\tilde \Delta_{SDW}(y)}{\tilde \Delta_{SDW}(0)} \sim (y_c-y)^{-2(G_3+G_2)d_1(y_c)}
.  \label{eq: SDW}
\end{eqnarray}
 This describes a potential instability towards 
  SDW formation, which will compete with the SCd instability. The SDW instability arises provided there is at least partial nesting i.e. the nesting parameter $d_1(y_c) \neq 0$. However, 
 since $-G_4>G_2$ for all $0\le d_1(y_c)\le 1$ (Fig.\ref{fig: rgintegration} inset), it follows from comparison of Eq.\ref{eq: SDW} and Eq.\ref{eq: exponents} that the SCd susceptibility diverges faster than the SDW susceptibility, for all values of nesting. At perfect nesting ($d_1=1$), the SCd susceptibility diverges as $(y_c-y)^{-1.5}$, whereas the SDW susceptibility diverges only as $(y_c-y)^{-1}$. As we move away from perfect nesting, the SCd susceptibility diverges faster, and the SDW susceptibility diverges more slowly, so that SCd is the leading instability for all values of nesting, within validity of the RG. 
This is in contrast to the square lattice \cite{Furukawa}, where  at perfect nesting the SDW and SCd instabilities have the same exponent under RG, with subleading terms lifting the degeneracy in favor of SDW, which is in turn overtaken by SCd at some $d_1<1$. 

We also considered 
the possibility of ordering in a channel exhibiting only a $\log^1$ divergence e.g. the Pomeranchuk ordering. However, we found that such orders cannot compete with superconductivity (see online supplementary material).
Finally, the phonon-mediated attraction in the pairing channel could induce $s$-wave superconductivity provided that it overwhelms the electronic repulsion in the s-wave channel at the Debye frequency scale, $\omega_D < \Lambda$.
%Finally, the electron-phonon coupling can in principle induce s-wave superconductivity. However, this will only occur if the phonon mediated attraction exceeds the electronic repulsion in the s-wave channel at the scale of the Debye frequency, $\omega_D < \Lambda$. 
However, the s-wave coupling ($2g(3)+g(4)$) remains positive and grows ever more repulsive under our $\log^2$ RG. Thus, as long as the nesting parameter is not too small ($2G_3>-G_4$ for $d_1>0.05$), the $s$-wave pairing appears to be unlikely to win.
%LL . Therefore, it seems unlikely that phonon mediated s-wave pairing can compete with d-wave superconductivity.}

\emph{Competition of d-wave orders below $T_c$:}
 We now investigate the competition of the $d_{x^2-y^2}$ and $d_{xy}$ superconducting orders (\ref{eq: deltas}) below $T_c$. In this regime, the system may either develop one of these two orders, or a linear combination of the two. 
The ordered state that minimizes the free energy wins. 
The hexagonal lattice point group symmetry dictates that the free energy below $
T_c$ must take the form \cite{Mineev} 
\bea
F  &=&\alpha (T-T_c) (|\Delta_a|^2 + |\Delta_b|^2)+ K_1 \big(|\Delta_a|^2 + |\Delta_b|^2\big)^2 
\nonumber \\
&& + K_2 |\Delta_a^2 + \Delta_b^2|^2
+ O(\Delta^6) 
\label{eq: landau}
\eea
with $K_1>0$. 
This free energy allows for two possible superconducting phases. If $K_2 <0$ then a $d_{x^2-y^2}$ or a $d_{xy}$ superconducting state would arise, whereas if if $K_2>0$ then the $d_{x^2-y^2}$ and $d_{xy}$ orders can co-exist \cite{Mineev}. We now calculate $K_2$ microscopically (an alternative but equivalent microscopic treatment is provided in the online supplementary material).

We begin by writing the free energy as the sum of the free energy on three patches, 
\begin{equation}
F = F(\Delta_1) + F(\Delta_2) + F(\Delta_3), \label{eq: sumf}
\end{equation}
 where the free energy on a patch is given by the standard Landau expansion
\begin{equation}
F(\Delta_i) = \alpha' (T - T_c) |\Delta_i|^2 + K |\Delta_i|^4, \quad K>0 \label{eq: onepatch}
\end{equation}
In this expression, it is essential to realize that $\Delta_1$, $\Delta_2$, and $\Delta_3$ are not independent, but must be expressed in terms of the two parameters $\Delta_{a}$ and $\Delta_{b}$, Eq.(\ref{eq: deltas}). Rewriting (\ref{eq: sumf}) and (\ref{eq: onepatch}) in terms of the two \emph{independent} variables $\Delta_{a,b}$, we obtain Eq.(\ref{eq: landau}) with $K_1 = \frac13 K>0$ and 
$K_2 = \frac16 K >0$. This implies the co-existence of $d_{x^2-y^2}$ and $d_{xy}$ orders. Minimization of the free energy (\ref{eq: landau}) with $K_2>0$ leads to $|\Delta_a|=|\Delta_b|$ and $\mathrm{Arg}(\Delta_a/\Delta_b) = \pi/2$. This order parameter can be rewritten as a three component vector in the patch basis, which takes the form
 \begin{equation}
\Delta_a \pm i \Delta_b = \Delta \big(1, e^{\pm 2\pi i/3}, e^{\mp 2\pi i/3} \big).
\end{equation}
 This corresponds to a superconducting gap that varies around the FS as $\Delta \exp(\pm 2i\phi)$.
 Such an order parameter corresponds to $d+id$ (or $d-id$) superconductivity (Fig.\ref{fig: orders}), and is a spin singlet analog of the $p+ip$ state that has been predicted for $Sr_2RuO_4$. 

%AC I would keep ``Conclusions''
{\it Conclusions:} The robustness of $d+id$ superconductivity in the weak coupling limit, demonstrated by the above analysis, leads us to believe that 
the graphene based chiral superconductivity can be realized experimentally. While our analysis is controlled for the weak short-range interaction model, several questions pertaining to the behavior of realistic systems should be clarified by future work. Determination of the phase structure for interactions of moderate strength and of long-range character remains an open problem, as does an accurate estimate of $T_c$ and the role of disorder, against which d-wave superconductivity is not protected. The graphene based $d+id$ superconductivity, if realized in experiment, will play a vital role in the development of technology designed to exploit topological superconductivity.
%In summary, we have shown that weak repulsive interactions induce superconductivity in graphene doped to the $M$ point. Using RG methods originally developed for the square lattice, we found that superconductivity unambiguously wins over other competing orders.  Furthermore, we have shown that the resulting superconducting state would constitute the first experimental realization of an exotic TRS breaking $d+id$ state, which is predicted to display an exceptionally rich phenomenology. This is particularly interesting given that the two-dimensional nature of graphene makes it exceptionally easy to construct heterostructures involving different types of materials. 
% , inducing isotropic BCS superconductivity, which may compete with $d+id$ pairing

% {\bf Author Contributions}: RN, LL and AC jointly identified the problem, performed the analysis and wrote the paper.

\section{Supplement}

\subsection{Estimating $T_c$}
We start by noting the tight binding dispersion
\begin{equation}
E = \pm t \sqrt{1+4 \cos(\pi k_x a)\big(\cos(\pi k_x a) + \cos(\sqrt3 \pi k_y a) \big)}
\end{equation}
where $t  =3 eV$ is the nearest neighbor hopping for graphene and $a \approx 1 \AA$ is the lattice constant. This defines a bandwidth $W\approx 9 ev$. %, the nearest neighbor hopping $t\approx 3 eV$, and the energy scale $\Lambda$ which is the UV cutoff for the RG. 
The linearization of the dispersion about the M point is valid for $\pi k a \ll 1$, which corresponds to an energy window of width $\Lambda_0 \approx t$. In this energy window, the density of states takes the form
\begin{equation}
\nu(E) \approx \nu_0 \ln \Lambda_0/E
\end{equation}
with $\nu_0 \sim 1/(2\pi t)$. 
%There is also an energy scale $\Lambda_0$ where the $k^4$ term in the dispersion becomes important. Since $E \sim t \big[cos(k_x) -cos(k_y)]$, it follows that $\Lambda_0 \sim t \sim 3 eV$. 

The RG is performed starting from some UV scale $\Lambda$. Since the RG is performed with the linearized dispersion, it is essential that $\Lambda \le \Lambda_0$. The natural choice would be to take $\Lambda = \Lambda_0 = t$. For technical reasons, however, it will prove convenient to take $\Lambda$ to be close to but slightly smaller than $\Lambda_0$. For the present, we simply treat $\Lambda$ as a free parameter. 

Now, we start with the Coulomb interaction $V(q) = 2\pi e^2/(q a^2)$ (we keep track of the lattice scale, so that the Coulomb interaction in momentum space continues to have dimensions of energy). Let us integrate out all states between the bandwidth W and the UV cutoff $\Lambda$. We will take into account the effect of these states in the RPA. The RPA can be formally justified by appealing to the large number of fermion flavors present in the problem, N=6 (3 patches and 2 spins). The effect of the high energy states in the RPA is to screen the Coulomb interaction
\begin{equation}
V(q,\omega) = \frac{\frac{2\pi e^2}{q a^2}}{1 + \frac{2 \pi e^2}{q a^2}N  \Pi_{\Lambda}(q,\omega)}
\end{equation}
where $N=6$ is the number of fermion flavors participating in screening and $\Pi_{\Lambda}$ is the single species polarisation function obtained by integrating out states with energies greater than $\Lambda$. The interactions $g_1$ and $g_4$ do not alter the patch label of electrons - treat these as having momentum transfer $q = q_{\Lambda} \approx 1/(\pi a)$, where $t q^2_{\Lambda}a^2 \approx \Lambda$. Meanwhile, the interactions $g_2$ and $g_3$ transfer momentum $Q = 1/a$, equal to the nesting vector connecting in-equivalent patches. Neglecting any frequency dependence of the interaction, 
\begin{eqnarray}
g_1(0)\approx g_4(0) \approx V(q_{\Lambda}, 0);\\%\approx \frac{2\pi^2 \nu_0  e^2/a }{1 + 2\pi^2 \nu_0 \ln(\Lambda_0/\Lambda) e^2/a}%= \frac{\frac{2\pi e^2}{q_{\Lambda}}{1+ \frac{2\pi e^2}{q_{\lambda}} \\
g_2(0) \approx g_3(0) \approx V(Q,0) %= \frac{2\pi e^2}{Q + (\epsilon-1)}.
\end{eqnarray}

Now, $\Pi_{\Lambda}$ should be like the polarisation function for an insulator with bandgap $\Lambda$. In particular, for large momentum transfer $q \ge q_{\Lambda}$, there should be metallic type screening, with $\Pi_{\Lambda} (q \ge q_{\Lambda}) \approx \nu(\Lambda) = \nu_0 \ln (\Lambda_0/\Lambda)$. Thus, we get
\begin{eqnarray}
g_1\nu_0  &=& g_4 \nu_0= \frac{\pi e^2/(at)}{1 + N \pi \ln (\Lambda_0/\Lambda) e^2/(at)}\\
g_1 \nu_0 &=& g_4 \nu_0= \frac{e^2/(at)}{1 + N \ln (\Lambda_0/\Lambda) e^2/(at)}
\end{eqnarray}
where we have taken $\nu_0 = 1/2\pi t$. Now $e^2/(at) \approx 5$ for $a = 1\AA$ and $t = 3 eV$, so we are close to unitarity. In this limit, we have 
\begin{equation}
g_1 \nu_0 \approx g_2 \nu_0 \approx g_3\nu_0 \approx g_4 \nu_0 \approx g_0 \nu_0= \frac{1}{N \ln (\Lambda_0/\Lambda)}
\end{equation}
Let us take $\ln \Lambda_0/\Lambda = 1$, i.e. $\Lambda = 1 eV$. There is admittedly an arbitrariness in this choice. However, this arbitrariness reflects itself only in an O(1) uncertainty in the prefactor for our eventual expression for $T_c$. Calculation of this prefactor is beyond the scope of RG, which can only calculate the exponent in the expression for $T_c$ (logarithmic accuracy). Thus, with logarithmic accuracy, we have $g_i = g_0 = 1/6$. We take these to be the bare couplings at the UV scale $\Lambda = 1eV$, and henceforth do RG. Substituting into the results obtained by integrating the $\ln^2$ RG equations, we obtain an estimate
% the result 
%
\begin{equation}
T_c = \Lambda \exp(-1.5/\sqrt{g_0 \nu_0}) \approx 200K
\end{equation}
(up to pre-factors of order unity). If true, this would be a remarkable result, exceeding the critical temperature of all other known superconductors. However, we have made some strong approximations in obtaining this result. For example, we have completely neglected single log terms in the RG equations. %This corresponds to neglect of e.g. screening arising from states with energies less than $\Lambda$. 
While such single log terms are formally subleading in the limit of weak coupling, they may well affect $T_c$. Moreover, since $T_c$ is exponentially sensitive to $g_0$, our approximate calculation of $g_0$ carries an exponentially large uncertainty in the value of $T_c$.

\subsection{The fixed point trajectory}

Here, we address the question of how large is the basin of attraction for the fixed point investigated in the main text. We show that the basin of attraction for the fixed trajectory includes the entire parameter space of weak repulsive interactions. We recall the RG equations
\begin{eqnarray}
\frac{dg_1}{dy} &=& 2 d_1 g_1 (g_2-g_1),  \qquad 
\frac{dg_2}{dy} = d_1(g_2^2 + g_3^2),\nonumber\\
\frac{dg_3}{dy} &=& - (n-2) g_3^2 - 2 g_3g_4 + 2 d_1g_3(2g_2-g_1), \qquad
\label{eq: beta functions supplement}\\\nonumber
\frac{dg_4}{dy} &=& -(n-1)g_3^2 - g_4^2 
\end{eqnarray}
We note 
%AC 
% in this regard 
%
that the equations (\ref{eq: beta functions supplement}) are homogenous, and the $\beta$ function for $g_2$ is positive definite. If we assume that the initial $g_2$ interaction is positive (repulsive), then $g_2$ is monotonically increasing under RG, and can be treated as a proxy for the RG time, following \cite{Vafek}. Making the substitutions $g_1 = x_1 g_2$, $g_3 = x_3 g_2$ and $g_4 = x_4 g_2$, we can rewrite (\ref{eq: beta functions supplement}) for $n=3$ as
\bea\nonumber
&&\frac{d x_1}{d \ln g_2} = - x_1 + \frac{2 x_1 (1-x_1)}{1+x_3^2},\quad 
\\
&&\frac{d x_3}{d \ln g_2} = - x_3 + \frac{2 d_1 x_3 (2-x_1) - x_3^2 -2x_3x_4}{d_1(1+x_3^2)}, \quad 
\label{eq: ratios}\\\nonumber
&&\frac{d x_4}{d \ln g_2} = - x_4 - \frac{2 x_3^2 + x_4^2}{d_1(1+x_3^2)} 
\eea
The fixed points of (\ref{eq: ratios})
%AC
 (e.g., solutions with constant $x_1$, $x_3$, and $x_4$) 
 correspond to fixed trajectories of the RG flow. When solving (\ref{eq: ratios})
 %AC 
 with $d x_i/d \ln g_2 =0$,
 $d_1$ should be interpreted as $d_1(y_c)$, and we should restrict ourselves to solutions with real values of $x_1$, $x_3$ and $x_4$, with $x_1>0$ and $x_3>0$. The latter constraint follows because the $\beta$ functions for $g_1$ and $g_3$ (\ref{eq: beta functions supplement}) vanish when the respective couplings go to zero, and so $g_1$ and $g_3$ cannot become negative if they start out positive.

%AC new
The set of non-linear algebraic equations for $x_i$ reduces to 7th order equation on. say, $x_1$, hence in general there are 7 different fixed trajectory. 
 However, three of them correspond to negative values of either $x_1$ or $x_3$,
 and three fixed trajectories are unstable, as we verified by solving the set (\ref{eq: ratios}) near the fixed trajectory. This leaves  
%AC  The
 the
 fixed trajectory discussed in the main text as the only stable fixed point of (\ref{eq: ratios}) that is compatible with the above constraints. Thus, any choice of weak repulsive interactions leads to the same fixed trajectory. 

%AC 
The solutions for $x_i$ along the fixed trajectory can be obtained analytically if we assume that the bare value of the exchange coupling $g_1$ is zero, in which case $g_1 =0$ holds during RG flow, and $x_1 =0$.  The set of two algebraic equations for $x_3$ and $x_4$ at the fixed point then reduces to 4th order polynomial algebraic equation, which can be solved exactly. Out of 4 soltions, two correspond to imaginary $x_3$ and one to a negative $g_3$. This leaves only one fixed trajectory, consistent with intial conditions.

\subsection{Ordering in $O(\ln)$ divergent channels}
Here we consider the possibility of ordering in an $O(\ln)$ divergent channel, and show that it cannot compete with superconductivity. First, we recall the scaling form of the superconducting susceptibility, 
\begin{equation}
\frac{\tilde \Delta_{a,b}(y)}{\tilde \Delta_{a,b}(0)} = \chi_{SCd}(y) \sim (y_c-y)^{2(G_4-G_3)} \sim \bigg(\frac{-g_0}{g_4(y)}\bigg)^{2(G_4-G_3)}, \label{eq: exponentssupplement}
\end{equation}

We wish to contrast this with the susceptibility in an $O(\ln)$ divergent channel. We therefore introduce a vertex corresponding to particle-hole pairing on the same patch, and examine how it renormalizes under RG. We find a scaling solution for the susceptibility, generic for any ordering in a $O(\ln)$ divergent channel, which takes the form  %which involves particle hole pairing at zero momentum. The flow equation for the susceptibility in a $O(\ln)$ divergent channel takes the form 
%
%\begin{equation}
%\frac{\partial \ln \chi_{DW}}{\partial y} = - \sum_{i=1}^4 \frac{b_i g_i(y)}{\sqrt{y}} \approx - \sum_{i=1}^4 b_i \frac{G_i }{\sqrt{y_c}(y_c-y)} .
%\end{equation}
%
%Integration of this equation leads to the result
%
\begin{equation}
\chi \sim (y_c-y)^{\alpha/\sqrt{y_c}} \sim \big(g_0/g_4(y) \big)^{\alpha \sqrt{g_0}}, \label{eq: chiln1}%\quad \alpha =  \sum_i b_i G_i 
\end{equation}
where $\alpha$ is some linear combination of the $G_i$ with $O(1)$ coefficients. Naively, such susceptibilities will also diverge as $y\rightarrow y_c$ if $\alpha<0$, although the exponent will be parametrically smaller than (\ref{eq: exponentssupplement}) by $\sqrt{g_0}$. However, we argue that not only is the exponent for these divergences parametrically small, but in fact such divergences lie outside the range of justifiable applicability of the RG. To understand why, it is essential to remember that the one loop RG only applies upto an energy scale $y_1$ when the couplings become of order one. (The limiting scale $y_1$ may actually be even smaller once we take into account self energy 
%AC 
$\Sigma (\omega, k_F) \propto g^2 \omega \log^2 \Lambda/|\omega|$ (Ref.\cite{self-energy})).

At the scale $y_1$, (\ref{eq: chiln1}) gives $\chi(y_1) = \exp\big(\alpha \sqrt{g_0} \ln g_0 \big)
$.% We now recall that $(y_c - y) \sim g_0/g(y)$ (where g is some diverging coupling). This allows us to rewrite (\ref{eq: scsusceptibility}) as %
%\begin{equation}
%\chi_{SCd}(y) \sim \bigg(\frac{g_0}{g(y)} \bigg)^{\alpha_{SCd}} \Rightarrow \chi_{SCd}(y_1) \sim g_0^{\alpha_{SCd}}
%\end{equation}
%where $\alpha_{SCd}$ is a negative constant. In the weak coupling limit $g_0 \rightarrow 0$ and $\chi_{SCd}(y_1) \rightarrow \infty$, indicating that a strong enhancement of the SCd susceptibility can occur within the realm of applicability of the one loop RG. Similar conclusions apply for the SDW instability. 
%In contrast, for pairing in $O(\ln)$ divergent channels, we can use the relation $y_c \sim 1/g_0$ to rewrite (\ref{eq: chiln1}) as
%\begin{equation}
%\chi(y) \sim \bigg(\frac{g_0}{g(y)} \bigg)^{\alpha \sqrt{g_0}}, \Rightarrow \end{equation}
In the weak coupling limit, $g_0 \rightarrow 0$ and $\chi(y_1) \approx 1$. Therefore, the susceptibility in a $\ln^1$ divergent channel is not significantly enhanced within the region of applicability of the one loop RG. In contrast, for $\ln^2$ divergent channels like SCd, (\ref{eq: exponentssupplement}) gives $ \chi_{SCd}(y_1) \sim g_0^{\alpha_{SCd}}$, which goes to infinity as $g_0 \rightarrow 0$. Therefore, only susceptibilities in $\ln^2$ divergent channels are strongly enhanced in the regime of justifiable applicability of weak coupling RG.

\subsection{Hubbard-Stratonovich treatment of superconductivity}

Here, we provide details of the Hubbard-Stratonovich treatment used to investigate the superconducting phase at temperatures lower than $T_c$. We begin by writing the partition function in the path integral formalism as $Z = \int D[\bar{\psi},\psi] \exp(-\int \mathcal{L}[\bar{\psi},\psi])$, where 
\bea
%Z &=& \exp \left(-\int \mathcal{L}\right) \nonumber\\
\mathcal{L} =&& \sum_{\alpha} \psi^{\dag}_{\alpha} (\partial_{\tau} - \epsilon_{\vec{k}} + \mu) \psi_{\alpha} \label{eq: scaction supplement}
\\ &-& \frac12 \left[ \begin{array}{c} \psi^{\dag}_1 \psi^{\dag}_1 \\ \psi^{\dag}_2 \psi^{\dag}_2 \\ \psi^{\dag}_3 \psi^{\dag}_3 \end{array}\right]^{\rm T} \left[\begin{array}{ccc} g_4 & g_3 & g_3 \\ g_3 & g_4 & g_3 \\ g_3 & g_3 & g_4\end{array} \right]\left[ \begin{array}{c} \psi_1 \psi_1 \\ \psi_2 \psi_2 \\ \psi_3 \psi_3 \end{array}\right] 
.
\eea
Here $\alpha$ is a patch index and the momentum, frequency and spin indices have been suppressed for clarity. We have included only the `pair hopping' interactions $g_3$ and $g_4$ since these are the only interactions that contribute to d-wave superconductivity.

As discussed above,
% We demonstrated earlier that under the 3-patch RG, $g_3$ and $g_4$ flow such that 
$g_3$ and $g_4$ flow under the 3-patch RG so that 
\begin{equation}
g_3 - g_4 = \lambda >0. 
\end{equation} 
When this is the case, then the interaction matrix in Eq.(\ref{eq: scaction supplement}) has two eigenvectors with degenerate negative eigenvalues. These reflect the two possible d-wave superconducting phases, which have identical instability threshold. We introduce two $3\times3$ matrices in patch space, $d_1$ and $d_2$, where 
\begin{equation}
d_a = \frac{1}{\sqrt{2}} \diag(0,1,-1); \quad d_b = \sqrt{\frac{2}{3}} \diag(1,-\frac12,-\frac12)
\end{equation}
These matrices obey $\Tr(d_a^2) = 1$, $\Tr(d_b^2)=1$ and $\Tr(d_a d_b)=0$, where the trace goes over the patch space. Using these matrices, we can define the order parameters of the two d-wave instabilities as $\Delta_a = 2 \lambda \langle \psi d_a \psi \rangle$ and $\Delta_b = 2 \lambda \langle \psi d_b \psi \rangle$. We can now decouple the quartic interaction in Eq.\ref{eq: scaction supplement} using a Hubbard Stratonovich transformation, and can hence rewrite the partition function as $Z = \int D[\bar{\psi},\psi, \Delta_{a,b}, \Delta_{a,b}^*] \exp(-\int \mathcal{L'}[\bar{\psi},\psi, \Delta_1, \Delta_1^*, \Delta_2, \Delta_2^*])$, where 
\bea
%Z &=& \exp \left(-\int \mathcal{L}\right) \nonumber\\
\mathcal{L'} = && \left[\begin{array}{c} \bar{\psi}_{\alpha} \\ \psi_{\beta}\end{array}\right]^{\rm T} \left[\begin{array}{cc} G_+^{-1} & \Delta_a d_a + \Delta_b d_b \\ \Delta_a^* d_a + \Delta_b^* d_b & G_-^{-1} \end{array}\right] \left[\begin{array}{c} \psi_{\alpha} \\ \bar \psi_{\beta} \end{array} \right] 
\nonumber\\
&&+  \frac{|\Delta_a|^2 + |\Delta_b|^2}{4\lambda}  \label{eq: hs}
\eea
We have written the action in a Gorkov-Nambu spinor form, introducing the particle and hole Green functions $G_+$ and $G_-$. These Green functions are diagonal in Fourier space, and have the form $G_{\pm}^{-1}(\omega, \vec{k}) = i\omega \mp (\epsilon_{\vec{k}} - \mu)$ 
 where $\omega$ is a Matsubara frequency, $\epsilon_{\vec{k}}$ is the energy of a state with momentum $\vec{k}$ and $\mu$ is the chemical potential. 
 We now integrate out the fermions in Eq.\ref{eq: hs} to obtain an exact action in terms of the order parameter fields alone. This action $\mathcal{L''}(\Delta_a,\Delta_a^*, \Delta_b, \Delta_b^*)$ takes the form
 \begin{equation}
 \mathcal{L''} = \Tr \ln \left( \begin{array}{cc} G_+^{-1} & \Delta_a d_a + \Delta_b d_b \\ \Delta_a^* d_a +\Delta_b^* d_b & G_-^{-1} \end{array} \right)
+  \frac{|\Delta_a|^2 + |\Delta_b|^2}{4\lambda}  \label{eq: hs_1}
 \end{equation}
The trace goes over patch space and over the spinor space. We expand this action in small $\Delta_{a,b}$ to order $\Delta^4$, exploiting the fact that the Green functions commute with the order parameter matrices, and the trace over patch space vanishes for any expression with an odd number of $d_a$ or $d_b$ matrices. We make use of the identities $\Tr(d_a^2) = \Tr(d_b^2) = 1$, $\Tr(d_a^4) = \Tr(d_b^4) = 1/2$, $\Tr(d_a^2 d_b^2) = \Tr(d_ad_bd_ad_b) = 1/6$, 
%AC 
 transform from partition function to free energy  and 
%AC hence
 obtain, up to an overall factor,
\bea
%\mathcal{L}
&& F = (|\Delta_a|^2 + |\Delta_b|^2)\left(\frac{1}{4\lambda} + \Tr(G_+ G_-) \right) + 
\label{eq: ch_1 supplement}\\\nonumber
&&   + K \left( |\Delta_a|^4  + |\Delta_b|^4 + \frac43 |\Delta_a|^2|\Delta_b|^2 + \frac{\bar \Delta_a^2 \Delta_b^2 + \bar \Delta_b^2 \Delta_a^2}{3}\right),
\eea
where $K = \Tr(G_+ G_- G_+ G_-) >0$.

 Superconductivity sets in when the coefficient of the quadratic terms first becomes negative, which leads to 
%AC an estimate for $T_c$,
%
\begin{equation}
T_c \sim \Lambda' \exp(-1/\sqrt{\lambda}) \label{eq: tcsqrt supplement}.
\end{equation}
%
%The $\sqrt{\lambda}$ dependence agrees with the results of the RG. It should be noted that (\ref{eq: tcsqrt}) is a weak coupling result, and when $\lambda$ becomes large, the increase of $T_c$ will be cut by $\ln^1$ contributions to the vertex corrections and the self energy. %Note that $T_c$ is identical for the two order parameters $\Delta_{a,b}$. 

The nature of the superconducting phase below $T_c$ %The instability threshold for superconductivity is found by looking for the temperature when the co-efficient of the quadratic term goes to zero. Since $\Tr(G_+G_-) =- \ln^2(\Lambda/T)$, this leads to an estimate 
%\begin{equation}
%T_c = \Lambda \exp\left(-1/\sqrt{g_3-g_4}\right)
%\end{equation}
%Note that $T_c$ is the same for both d-wave order parameters. The form of the superconducting state below $T_c$ may be deduced from 
is controlled by the anisotropic quartic term.
%AC 
 Since $K>0$, minimization of the quartic term leads to $|\Delta_a|=|\Delta_b|$ and $\mathrm{Arg}(\Delta_a/\Delta_b) = \pi/2$. 
The full superconducting order parameter is thus 
\bea
\Delta_a \pm i \Delta_b &=& \Delta\sqrt{\frac{2}{3}}\lp 1,-\frac12\pm i\frac{\sqrt{3}}2,-\frac12\mp i\frac{\sqrt{3}}2\rp
\nonumber \\
&=&\Delta\sqrt{\frac{2}{3}}\lp 1,e^{\pm 2\pi i/3},e^{\mp 2\pi i/3}\rp.
% \Delta\bigg(\sqrt{\frac{2}{3}},\pm \frac{i}{\sqrt{2}} - \frac{1}{\sqrt6}, \mp \frac{i}{\sqrt2} - \frac{1}{\sqrt6} \bigg).
\eea

%Alternatively, the expansion of  the Free energy in powers of $\Delta_{1,2}$  can be obtained by One can easily make sure that this is the same expression as Eq. (\ref{ch_1}). 
%\end{document}

% \begin{thebibliography}{99}
% \vspace{-7mm}

% \providecommand{\natexlab}[1]{#1}
% \providecommand{\url}[1]{\texttt{#1}}
% \expandafter\ifx\csname urlstyle\endcsname\relax
%   \providecommand{\doi}[1]{doi: #1}\else
%   \providecommand{\doi}{doi: \begingroup \urlstyle{rm}\Url}\fi
% \bibitem[Vafek (2010)]{Vafek}
% Vafek, O. and Yang, K. Many-body instability of Coulomb interacting bilayer graphene. {\it Phys. Rev. B} {\bf 81}, 041401(R) (2010)

% \bibitem[selfenergy (2005)]{self-energy} see e.g., Katanin, A.A., Kampf, A.P. and Irkhin, V.Y. Anomalous self-energy and Fermi surface quasisplitting in the vicinity of a ferromagnetic instability. {\it Phys. Rev. B} {\bf 71}, 085105 (2005) and references therein. 

% \end{thebibliography}

\end{document}